\documentclass[graybox]{svmult}

\usepackage{mathptmx}       
\usepackage{helvet}         
\usepackage{courier}        
\usepackage{type1cm}        
\usepackage{makeidx}         
\usepackage{graphicx}        
\usepackage{multicol}        
\usepackage[bottom]{footmisc}

\usepackage[nooneline,raggedright]{subfigure}
\newcommand{\oo}[1]{\frac{1}{#1}}

\newcommand{\fref}[1]{Fig.~\ref{#1}}
\newcommand{\sref}[1]{Section~\ref{#1}}

\makeindex             

\begin{document}

\title*{Thermopower of the Correlated Narrow Gap Semiconductor FeSi
 and Comparison to RuSi}
 \titlerunning{Thermopower of the Correlated Narrow Gap Semiconductor FeSi}

\author{Jan M. Tomczak, K. Haule, G. Kotliar}
\institute{Jan M. Tomczak \at Department of Physics and Astronomy, Rutgers University, Piscataway, New Jersey 08854, USA \email{jtomczak@physics.rutgers.edu}
\and K. Haule \at Department of Physics and Astronomy, Rutgers University, Piscataway, New Jersey 08854, USA \email{haule@physics.rutgers.edu}
\and G. Kotliar \at Department of Physics and Astronomy, Rutgers University, Piscataway, New Jersey 08854, USA \email{kotliar@physics.rutgers.edu}}
%
%
\maketitle

\abstract{
Iron based narrow gap semiconductors such as FeSi, FeSb$_2$, or FeGa$_3$
have received a lot of attention because they exhibit a large thermopower, as well as
striking similarities to heavy fermion Kondo insulators.
Many proposals have been advanced, however, lacking quantitative methodologies applied to this problem, a consensus remained elusive to date.
Here, we employ realistic many-body calculations
to elucidate the impact of electronic correlation effects on FeSi.
Our methodology accounts for all substantial anomalies observed in FeSi: the metallization, the lack of conservation of spectral weight in optical spectroscopy, and the Curie susceptibility.
In particular we find a very good agreement for the anomalous thermoelectric power.
Validated by this congruence with experiment, we further discuss a new physical picture of the microscopic nature of the insulator-to-metal crossover.
Indeed, we find the suppression of the Seebeck coefficient to be driven by correlation induced incoherence.
Finally, we compare FeSi to its iso-structural and iso-electronic homologue RuSi, and
predict that partially substituted Fe$_{1-x}$Ru$_x$Si will exhibit an increased thermopower at intermediate temperatures. 
}

\section{Introduction}

Correlated semiconductors have been a subject of intensive research over the years, because they
exhibit an unusual metalization process which is poorly understood. At low temperatures, the iron silicide FeSi
-- the prototypical compound of this class of materials --
is akin to an ordinary semiconductor with a gap of $\Delta\approx50-60$meV
\cite{PhysRev.160.476,PhysRevLett.71.1748,0295-5075-28-5-008,PhysRevB.56.12916}. Yet, at higher temperatures, that are however much smaller than $\Delta/k_B$, FeSi becomes
a bad metal\cite{Wolfe1965449,PhysRevLett.71.1748,PhysRevB.56.12916} and develops a Curie-Weiss   
like susceptibility\cite{PhysRev.160.476,JPSJ.50.2539}. Analogies with heavy fermion Kondo insulators\cite{ki,Schlesinger1997460,PhysRevB.51.4763} and mixed valence compounds\cite{PhysRevB.50.9952}, effects of spin-fluctuations\cite{JPSJ.46.1451}, or spin-state transitions\cite{PhysRevLett.76.1735,vanderMarel1998138},
as well as an anomalous electron--phonon coupling\cite{PhysRevB.59.15002,PhysRevB.83.125209,Delaire22032011} have been invoked to account for this behaviour. 
Lacking, however, quantitative methodologies applied to this problem, a consensus remained elusive to date.

Besides this fundamental puzzle, the class of correlated narrow gap semiconductors, which also comprises compounds such as
FeSb$_2$, FeAs$_2$ or FeGa$_3$,
is also of interest in view of applied science. Indeed these materials exhibit notably large Seebeck coefficients at low temperatures\cite{Wolfe1965449,Buschinger1997784,0295-5075-80-1-17008, PhysRevB.83.125209,sun_dalton,APEX.2.091102,jmt_fesb2,JPSJ.78.013702}, and the largest thermoelectric
powerfactor ever to be measured was recently found for FeSb$_2$\cite{0295-5075-80-1-17008}. 
An understanding of the electronic structure of these compounds, and the potential influence of electronic correlation effects thereon
is thus of vital interest.

Here, we will investigate the paradigmatic example FeSi by means of the realistic extension of dynamical mean-field theory (see e.g.\ Ref.~\cite{RevModPhys.78.865} for a review).
Our recent results\cite{jmt_fesi} account for all substantial anomalies observed in FeSi, namely the lack of conservation
of spectral weight in the optical conductivity\cite{PhysRevLett.71.1748,0295-5075-28-5-008,PhysRevB.55.R4863,vanderMarel1998138,PhysRevB.56.12916,PhysRevB.79.165111}, a Curie-Weiss like susceptibility and an anomalous thermoelectric power. 
From the microscopic insight of our approach, we elucidate the origin
of the metal-insulator transition. We explain that the latter is a consequence of a correlation induced incoherence:
Unlike in conventional semiconductors where a metalization process is driven
by thermal activation or a moving of the chemical potential into the conduction or valence bands, in correlated insulators such as FeSi the crossover 
is induced by the emergence of non-quasiparticle incoherent states in the gap.
We further link the occurrence of these many-body states to the spin degrees of freedom.

With this understanding we here address the Seebeck coefficient of FeSi, and propose ways on how to improve the thermoelectric performance
of FeSi based systems. In particular, we compare FeSi to its isoelectronic homologue RuSi, and speculate on the thermoelectric properties of Ru substituted FeSi.

\begin{table}%
\begin{tabular}{cccc}
{``NaCl''} &  &  & {B20} \\
{ \includegraphics[angle=0,width=0.225\textwidth]{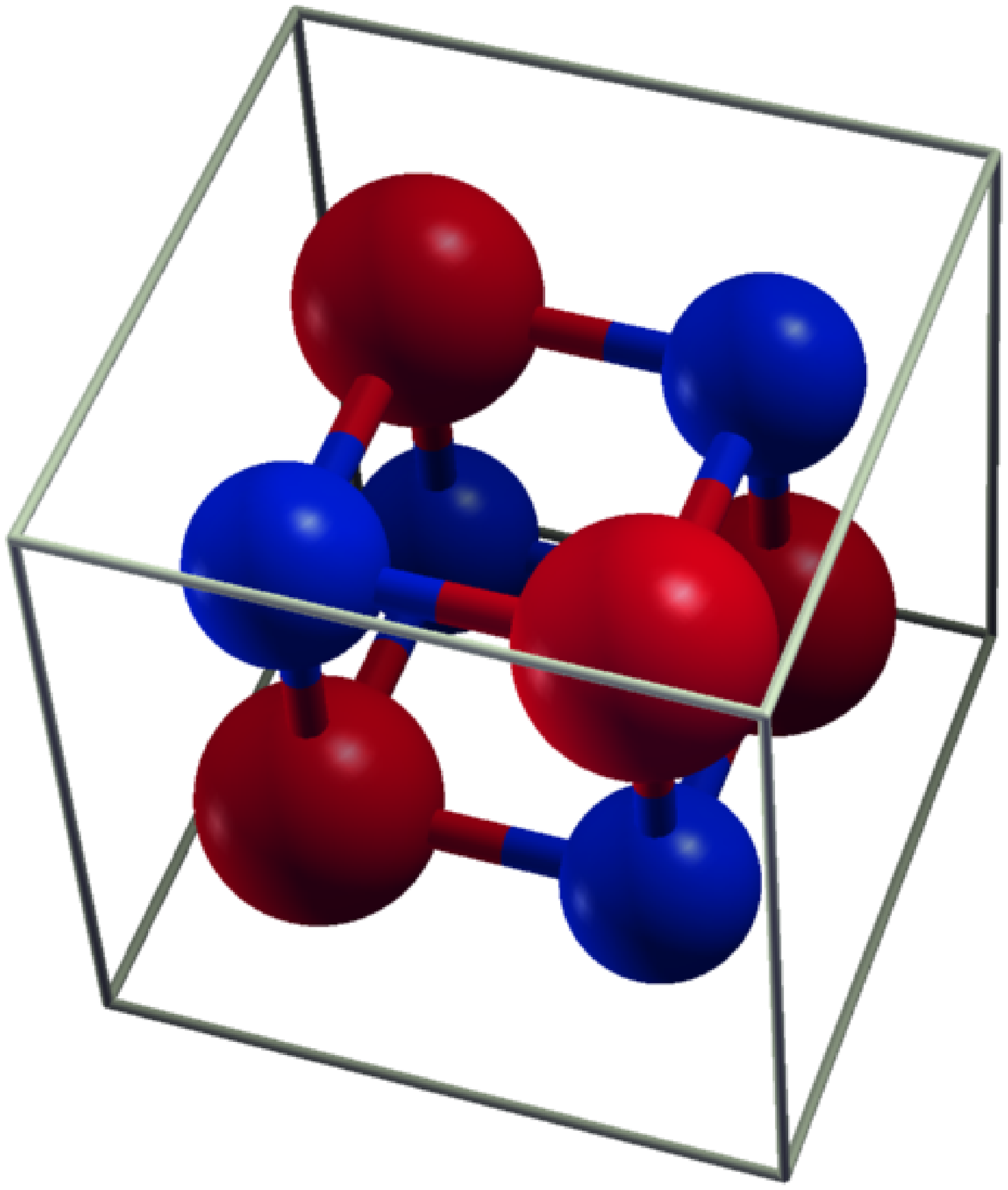}} &
{ \includegraphics[angle=0,width=0.225\textwidth]{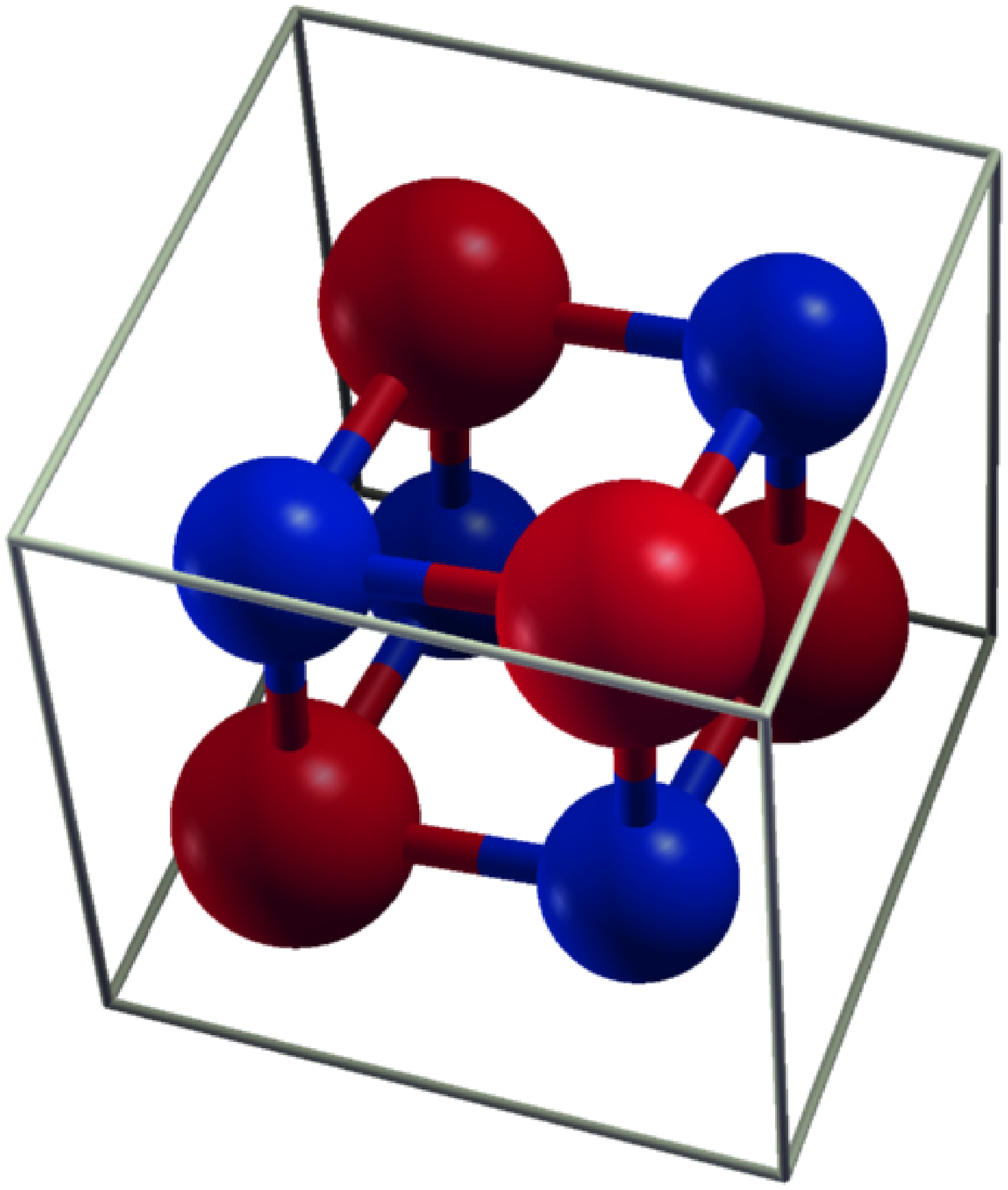}} &
{ \includegraphics[angle=0,width=0.225\textwidth]{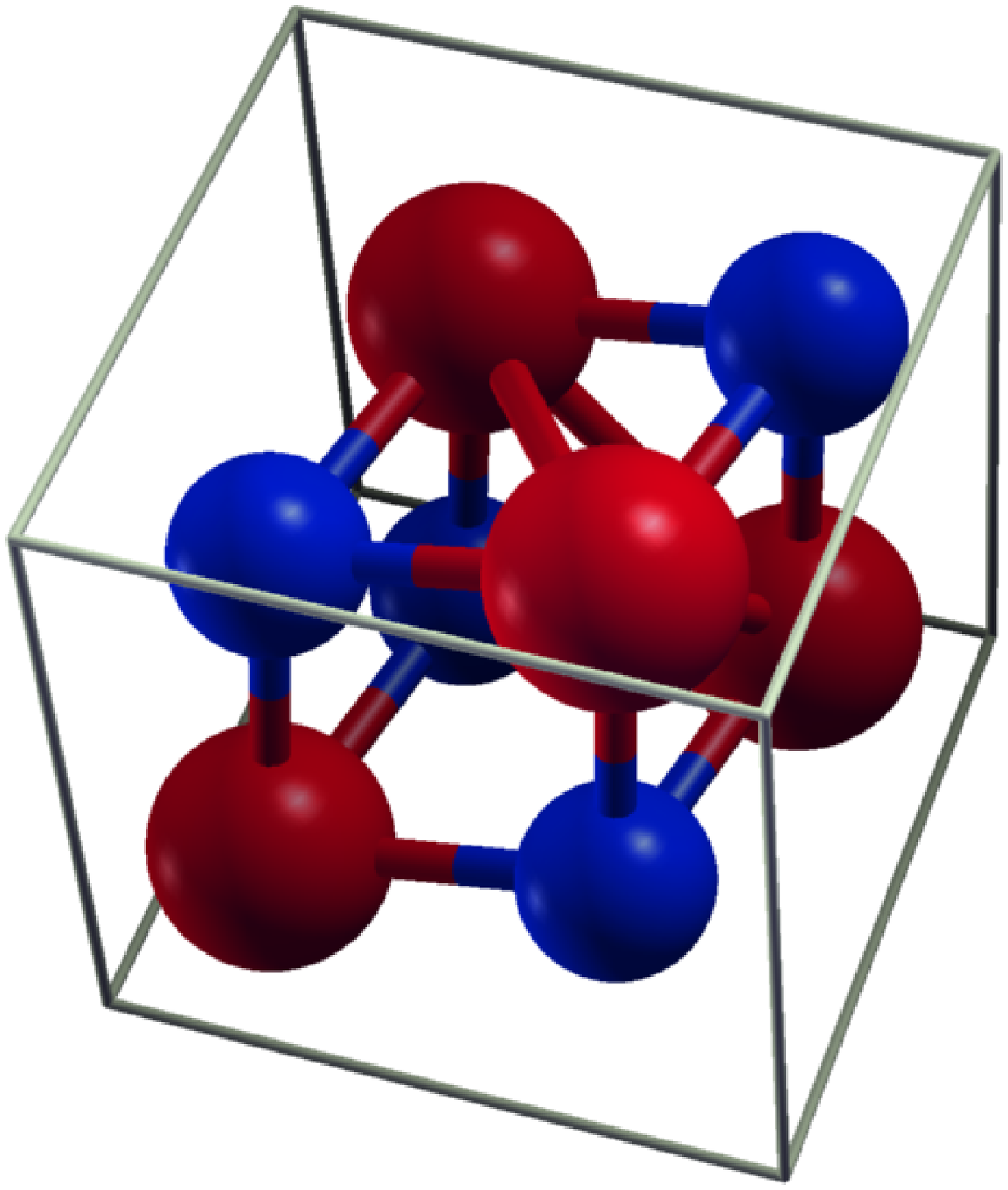}} &
{ \includegraphics[angle=0,width=0.225\textwidth]{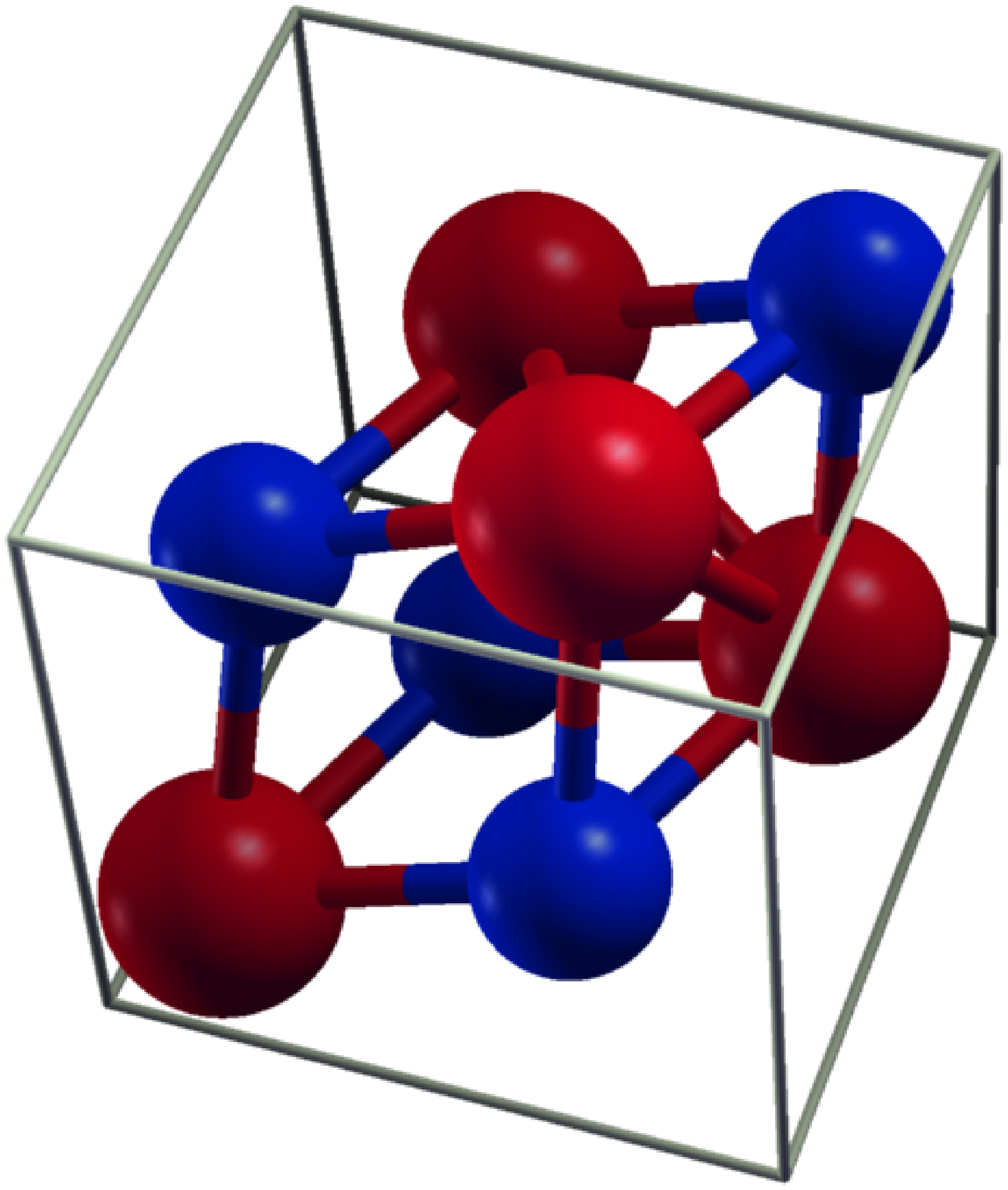}} \\
{ \includegraphics[clip=true,trim=0 50 20 20, angle=0,width=0.235\textwidth]{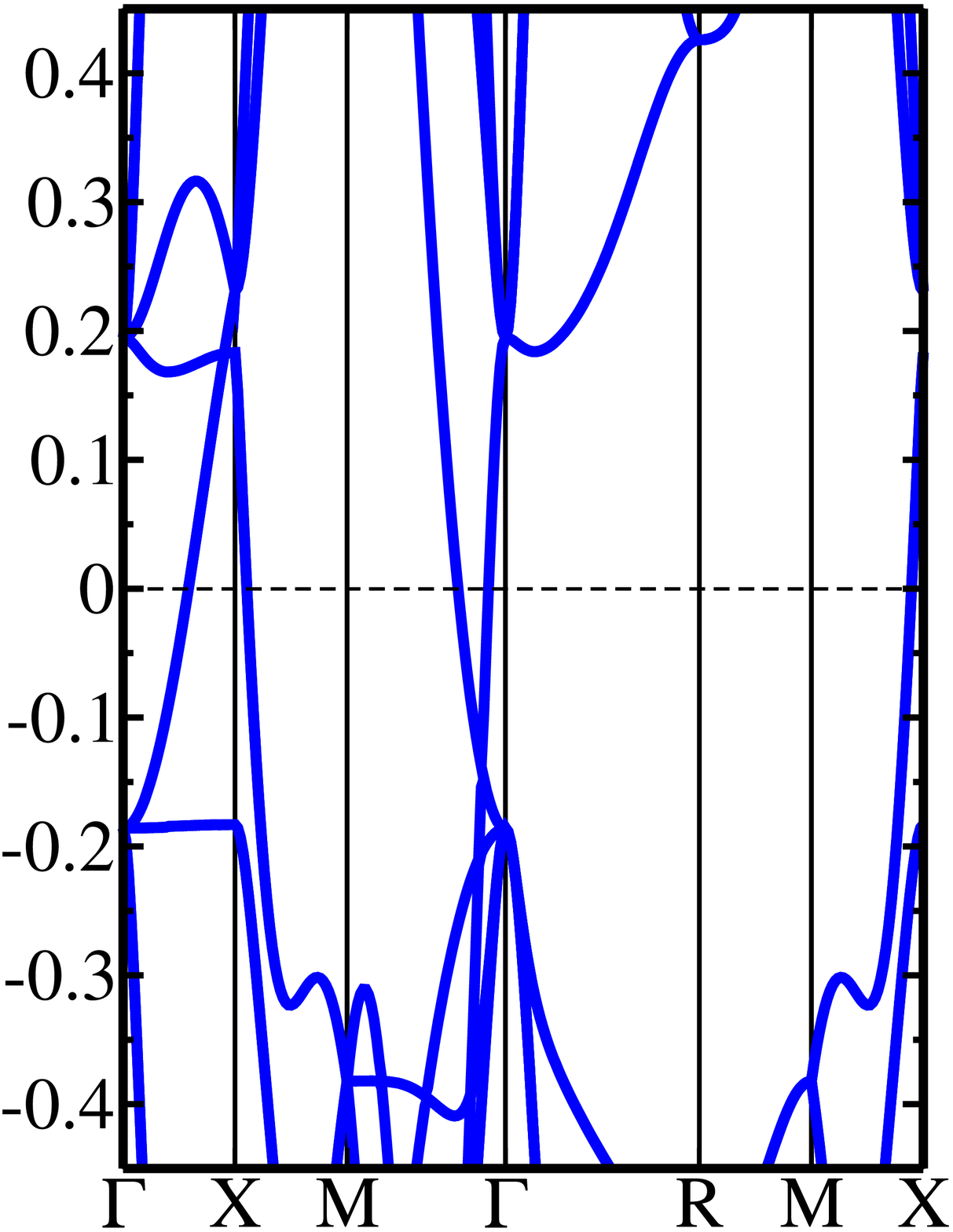}} &
{ \includegraphics[clip=true,trim=0 50 20 20, angle=0,width=0.235\textwidth]{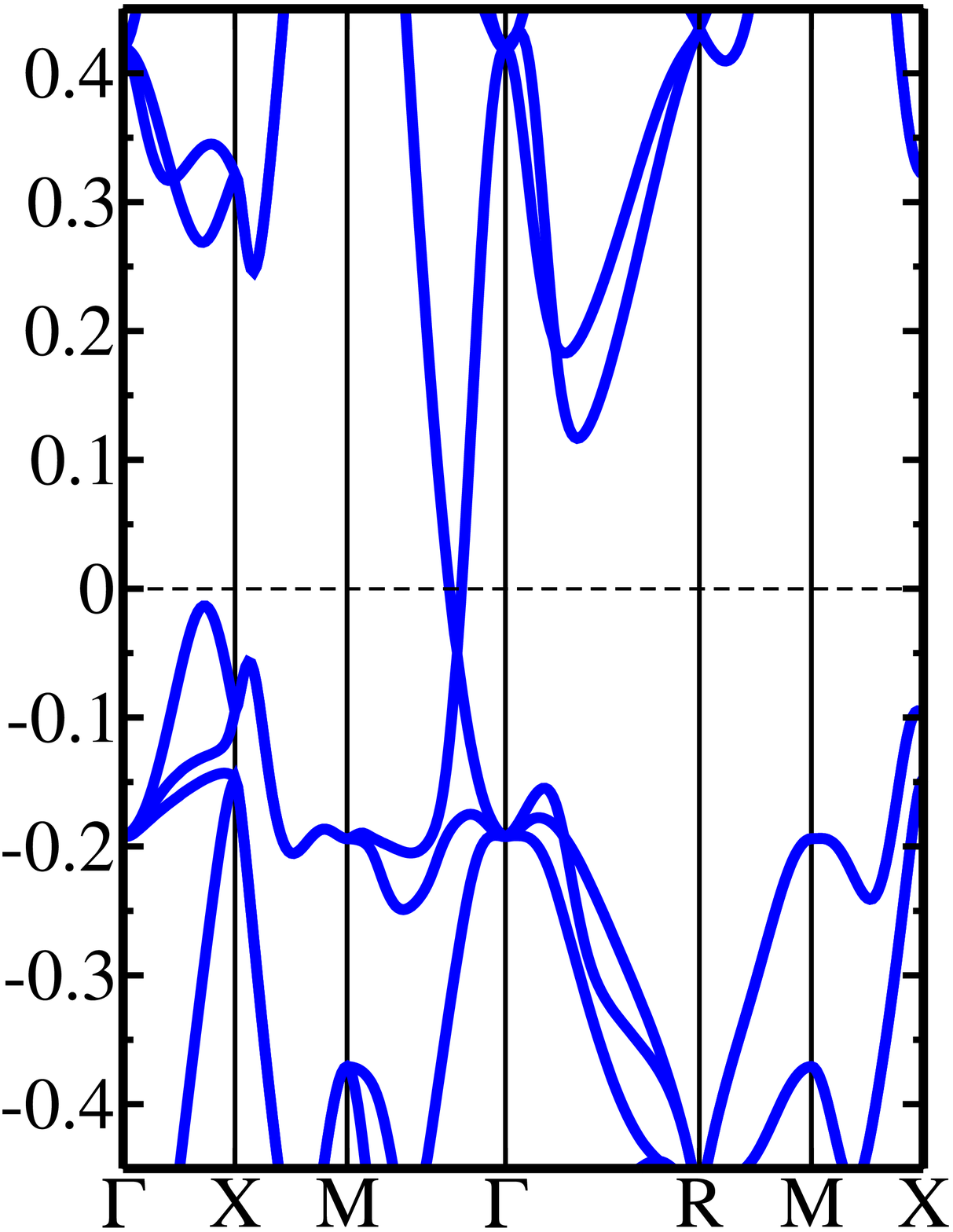}} &
{ \includegraphics[clip=true,trim=0 50 20 20, angle=0,width=0.235\textwidth]{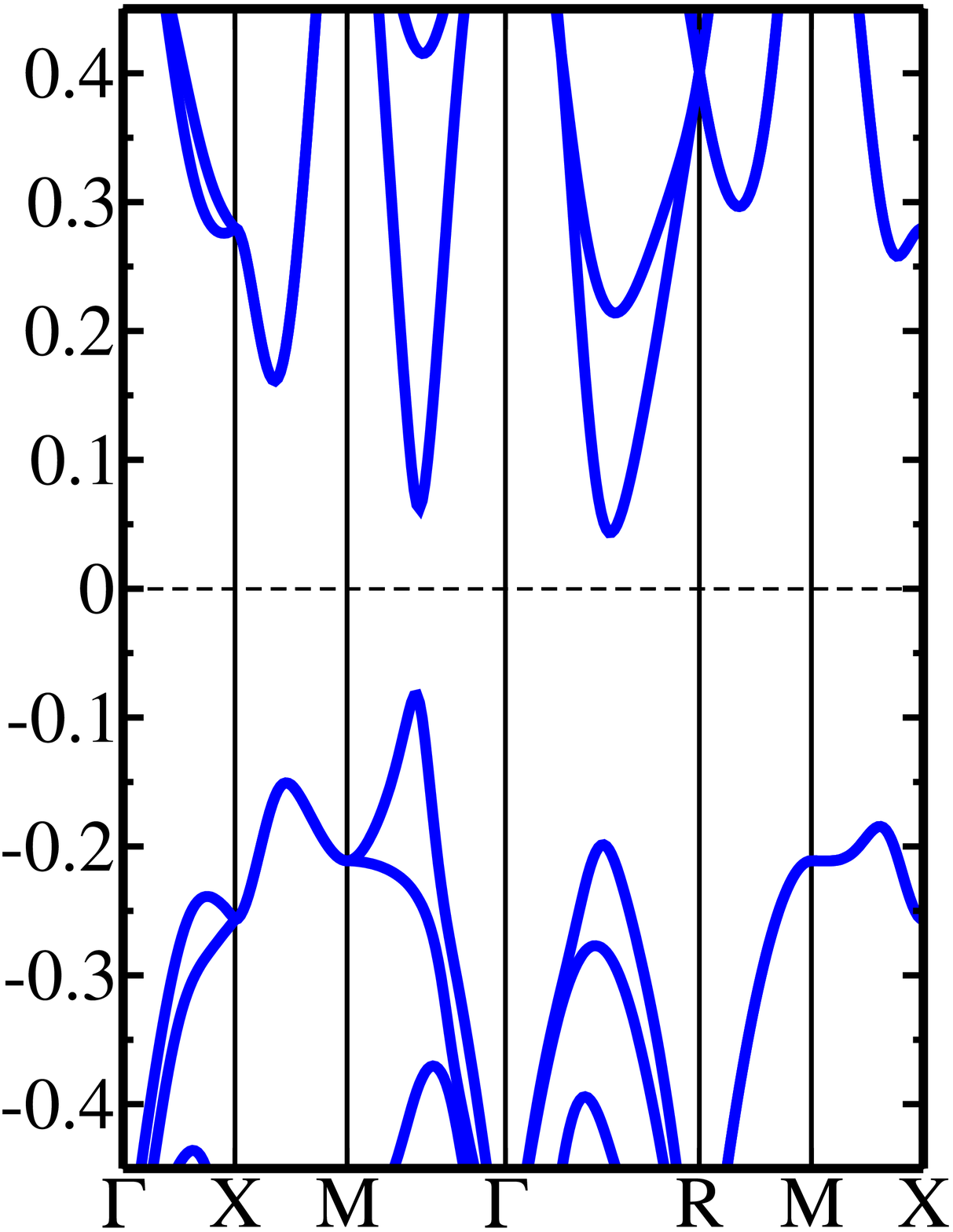}} &
{ \includegraphics[clip=true,trim=0 50 20 20, angle=0,width=0.235\textwidth]{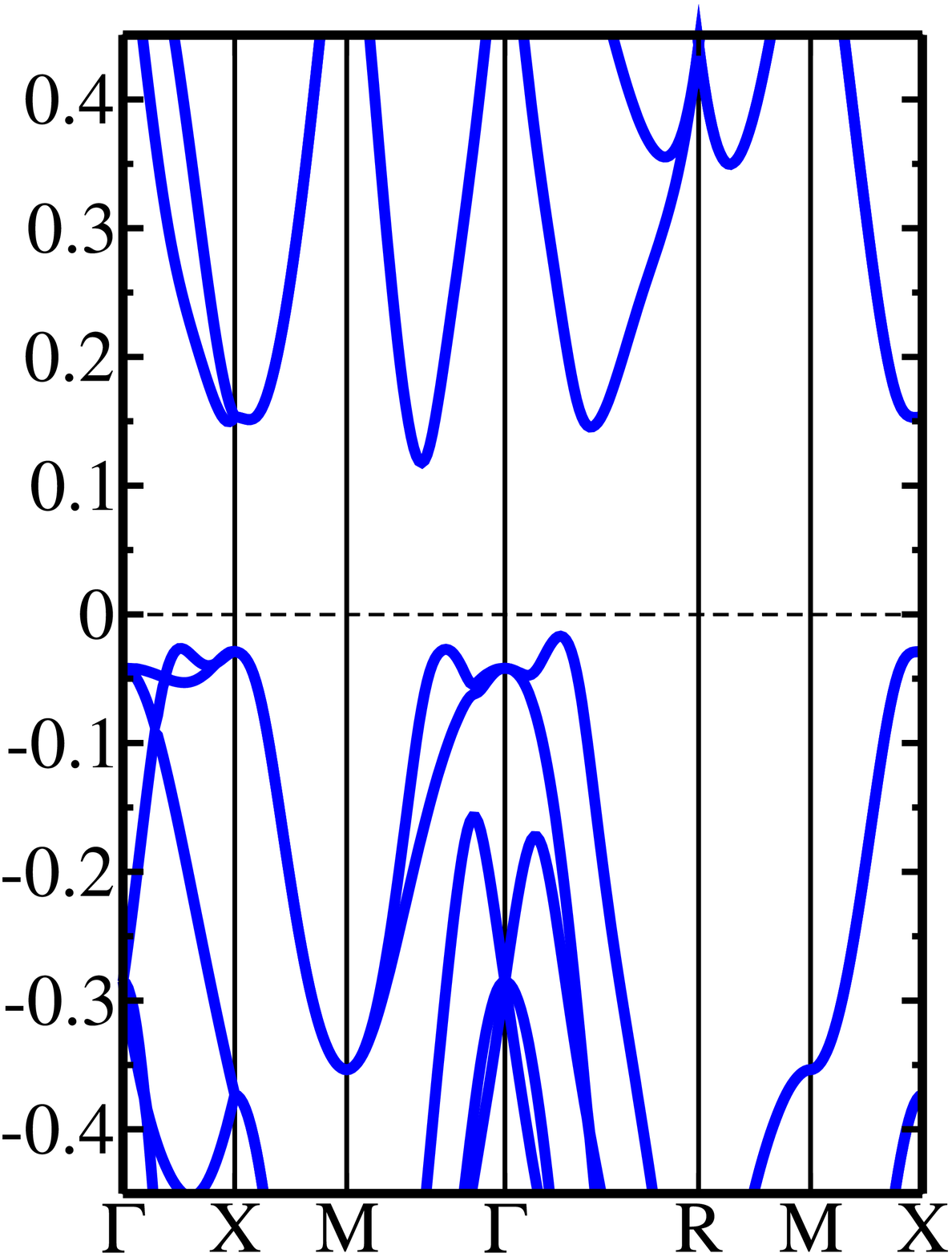}} 
\end{tabular}
\caption{Band-structure of FeSi in the B20 structure for different degrees of the distortion from the simple cubic rock-salt structure.
Energies in electronvolts, the origin corresponds to the Fermi level.
From left to right: ``NaCl'' rock-salt structure 0\%, 50\%, 75\%, and 100\% distortion corresponding to actual FeSi.
Above the band-structures are shown the respective unit cells (pictures made with \cite{Anton1999176}), iron atoms are red, silicon ones blue.}
\label{fig_bnd}
\end{table}

\section{The metal-insulator transition in FeSi and its microscopic origin}
\subsection{Crystal- and band-structure}
\label{bnd}
The B20 crystal structure of FeSi, despite being cubic (space group P2$_1$3), is rather complex and has four iron atoms per unit cell.
However, it can be viewed as a simple cubic rock-salt structure that is highly distorted along the [111] direction\cite{PhysRevB.47.13114}%
\footnote{Indeed,
the positions of the iron and silicon atoms are (u,u,u), ($\oo{2}$+u,$\oo{2}$-u,1-u), (1-u,$\oo{2}$+u,$\oo{2}$+u), and ($\oo{2}$-u,1-u,$\oo{2}$+u)
with u(Fe)=1/4 and u(Si)=3/4 for the rock-salt structure, and u(Fe)=0.136 and u(Si)=0.844 for FeSi\cite{PhysRevB.47.13114}.}.
In order to elucidate the important effect of this distortion onto the band-structure, we display in \fref{fig_bnd} the evolution
of the Kohn-Sham spectrum (using the generalized gradient approximation (GGA) functional within density functional theory (DFT) as implemented in wien2k\cite{wien2k}) for different degrees of the atomic displacements%
\footnote{as measured by a linear interpolation of the u parameters from above (see also Ref. \cite{PhysRevB.81.125131}).}.
While FeSi is metallic in the fictitious rock-salt structure, the changes in the atomic positions cause the formation of an avoided crossing in the distorted B20 structure. In Fesi, the iron 3d orbitals split into a low-lying $z^2$, and the two doubly degenerate groups $x^2$-$y^2$, $xy$ and $xz$, $yz$, which are separated by a band gap\cite{PhysRevB.47.13114,PhysRevB.59.15002,PhysRevB.81.125131,jmt_fesi}.
With a nominal valence of 6 electrons per iron, band-structure methods thus correctly reproduce the low temperature insulating character of FeSi.
Owing to the hybridization nature of the gap, as well as the absence of many-body band-narrowing effects within DFT methods,
 the gap is overestimated by a factor of two,
$\Delta_{DFT}\approx0.11$eV\cite{PhysRevB.47.13114,PhysRevB.59.15002,PhysRevB.81.125131}.

\begin{figure}[htpangle=-90]
\centering
\includegraphics[angle=-90,width=0.8\textwidth]{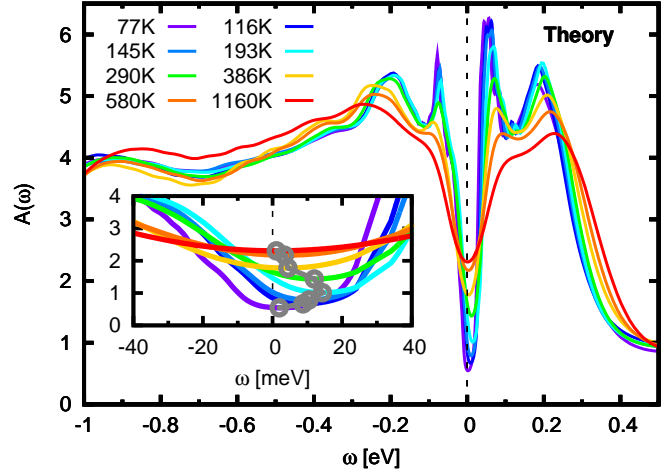}
\caption{{\bf Spectral function.} LDA+DMFT spectrum at various temperatures. The inset shows a zoom of the low-energy pseudo-gap, and grey symbols indicate the spectral minimum, which moves as a function of temperature owing to the particle/hole asymmetry of the spectrum.}
\label{fig_spec}
\end{figure}

\subsection{The many-body spectrum}

While the shown band-structure may qualitatively describe the low temperature properties of FeSi, e.g.\ the activated behaviour found in the resistivity below 150K\cite{Wolfe1965449,Buschinger1997784},
it obviously cannot explain the insulator to metal transition at higher temperatures, let alone a Curie-Weiss like tail in the spin susceptibility.

To account for electronic correlation effects beyond the effective one-particle description of band-theory, we resort to the realistic extension ``LDA+DMFT'' of dynamical mean-field theory\cite{RevModPhys.78.865},
as implemented in Ref.~\cite{PhysRevB.81.195107}.
We use established values of the local Coulomb interaction in iron based compounds \cite{PhysRevB.82.045105} (U=5.0eV, J=0.7eV), and solve the DMFT impurity by means of a continuous time quantum Monte Carlo (ctqmc) method\cite{PhysRevB.75.155113,PhysRevLett.97.076405}.

\begin{figure}[htp]
\centering
\includegraphics[angle=-90,width=0.8\textwidth]{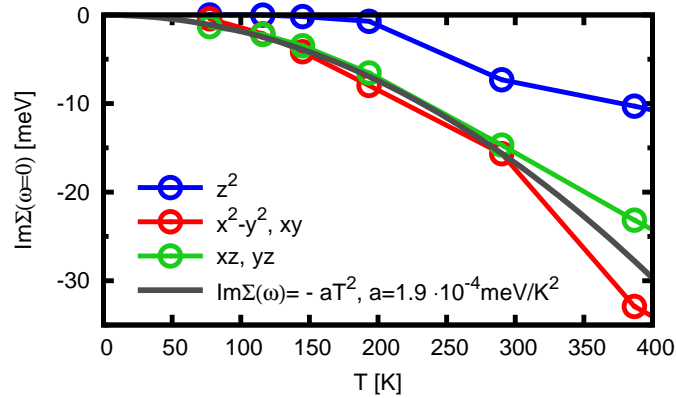}
\caption{{\bf Correlation induced incoherence.} Imaginary part of the DMFT self-energy at the Fermi level as a function of temperature and resolved into orbital characters. Also shown
is the quadratic fit: $\Im\Sigma(\omega=0)=-1.9\cdot 10^{-4}$meV/K$^2$ $\cdot$ T$^2$ for the $x^2$-$y^2$/$xy$ and $xz$/$yz$ components.}
\label{fig_ImS}
\end{figure}

In \fref{fig_spec} are displayed the local spectral functions that we obtain for different temperatures as indicated.
The low temperature spectrum is akin to the density of states obtained within band-theory, yet with a gap that is renormalized by about a factor of two, in agreement
with photoemission spectroscopy experiments\cite{PhysRevLett.101.046406,PhysRevB.77.205117}.
This means that lifetime effects are minor at low temperatures, as also inferred from the self-energy shown in \fref{fig_ImS}. 
In particular the spectral weight at the edges of the gap is remarkably spiky, a characteristic that is commonly considered a hallmark for a potentially large thermopower\cite{Mahan07231996}.
With increasing temperature however, spectral features broaden, and the charge gap gets filled with incoherent weight, with solely a pseudo-gap remaining.
This is in congruence with experimental findings, both in one-particle probes (such as photoemission\cite{PhysRevLett.101.046406,PhysRevB.77.205117}), transport measurements (for a comparison of our theoretical resistivity
with experiments, see Ref. \cite{jmt_fesi}), and optical spectroscopy (again, see Ref. \cite{jmt_fesi}).
Besides the filling of the gap, a further detail of photoemission results is captured:
Since the spectrum is particle/hole asymmetric, the chemical potential moves as a function of temperature.
As a measure for this, we mark in the inset of \fref{fig_spec} the position in energy at which spectral weight is minimal.
At low temperatures, the latter is in the vicinity of the chemical potential
(as expected for a semiconductor\cite{jmt_fesb2}). Since the unoccupied states have  larger spectral weight than the valance states, the chemical potential moves down upon increasing temperature,
causing the spectral minimum to move up in energy, as indeed found in photoemission\cite{PhysRevLett.101.046406,PhysRevB.77.205117}.
Above 300K, the asymmetry switches, and the position of the spectral minimum moves back towards the chemical potential.
These trends in the particle/hole symmetry are harbingers of the temperature dependence of the Seebeck coefficient, discussed in \sref{seebeck}.

Thus, as far as spectral properties are concerned our calculations are a realistic generalization of the seminal model of Fu and Doniach\cite{PhysRevB.51.17439}.
However, as we detail in Ref. \cite{jmt_fesi}, there is a fundamental physical difference between that model and our results.
While the degree of correlations in the two band model\cite{PhysRevB.51.17439} is controlled by Hubbard physics, we find our results
to be much more sensitive to the Hund's rule coupling $J$ than the Hubbard $U$.
This highlights the multiorbital nature of the system and, hence, the necessity of realistic calculations.
For a more detailed discussion, see Ref.\cite{jmt_fesi}. In the context of FeSi note also the recent Ref.\cite{0295-5075-95-4-47007}, as well as 
Refs. \cite{1367-2630-11-2-025021,PhysRevLett.106.096401,Yin_pnictide} for the influence of the Hund's rule coupling in other systems.

\subsection{The physical picture -- the self-energy and the spin state}

As seen above, the insulator to (bad) metal transition is not caused by a closure of the charge gap, but by a filling of the latter with incoherent spectral weight.
The degree of incoherence of the many-body system is encoded in the imaginary part of the electron self-energy, $\Im\Sigma(\omega)$. In \fref{fig_ImS} we depict the zero frequency limit
of that quantity, resolved into the iron 3d orbital characters. The orbital components that account for spectral weight at low energies are, as mentioned above, the
$x^2$-$y^2$ and $xy$ on the valence, and  the $xz$ and $yz$ on the conduction side. As seen in the figure, the latter two neatly follow a T$^2$ law for the temperature range shown,
with a coefficient of $-1.9\cdot 10^{-4}$meV/K$^2$. Therewith the inverse lifetime quickly becomes comparable to the size of the gap, which it surpasses at about 400K.

\begin{figure}[th]
  \begin{center}
\subfigure[\, ]{\includegraphics[angle=-90,width=0.49\textwidth]{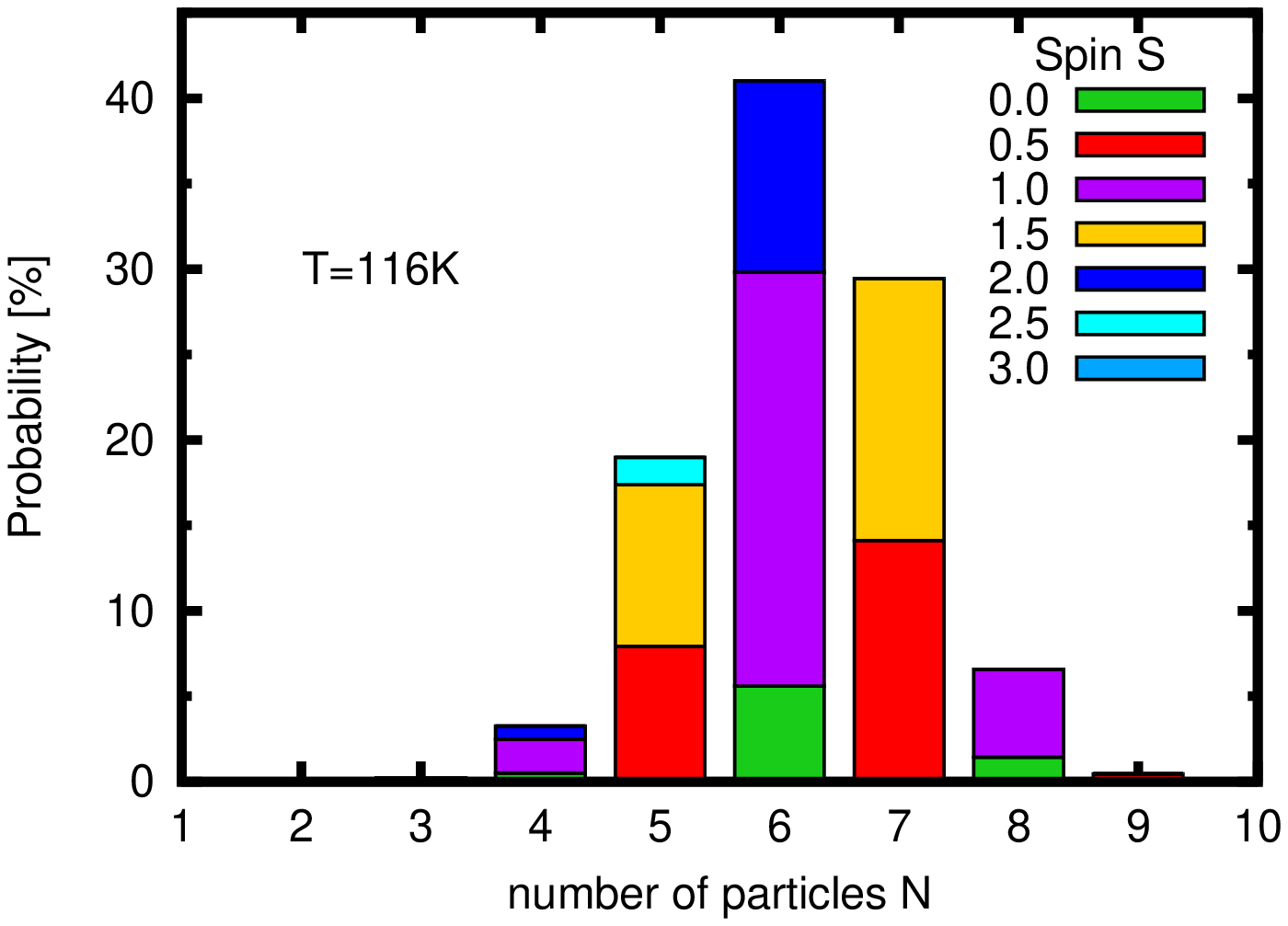}}
\subfigure[\, ]{\includegraphics[angle=-90,width=0.49\textwidth]{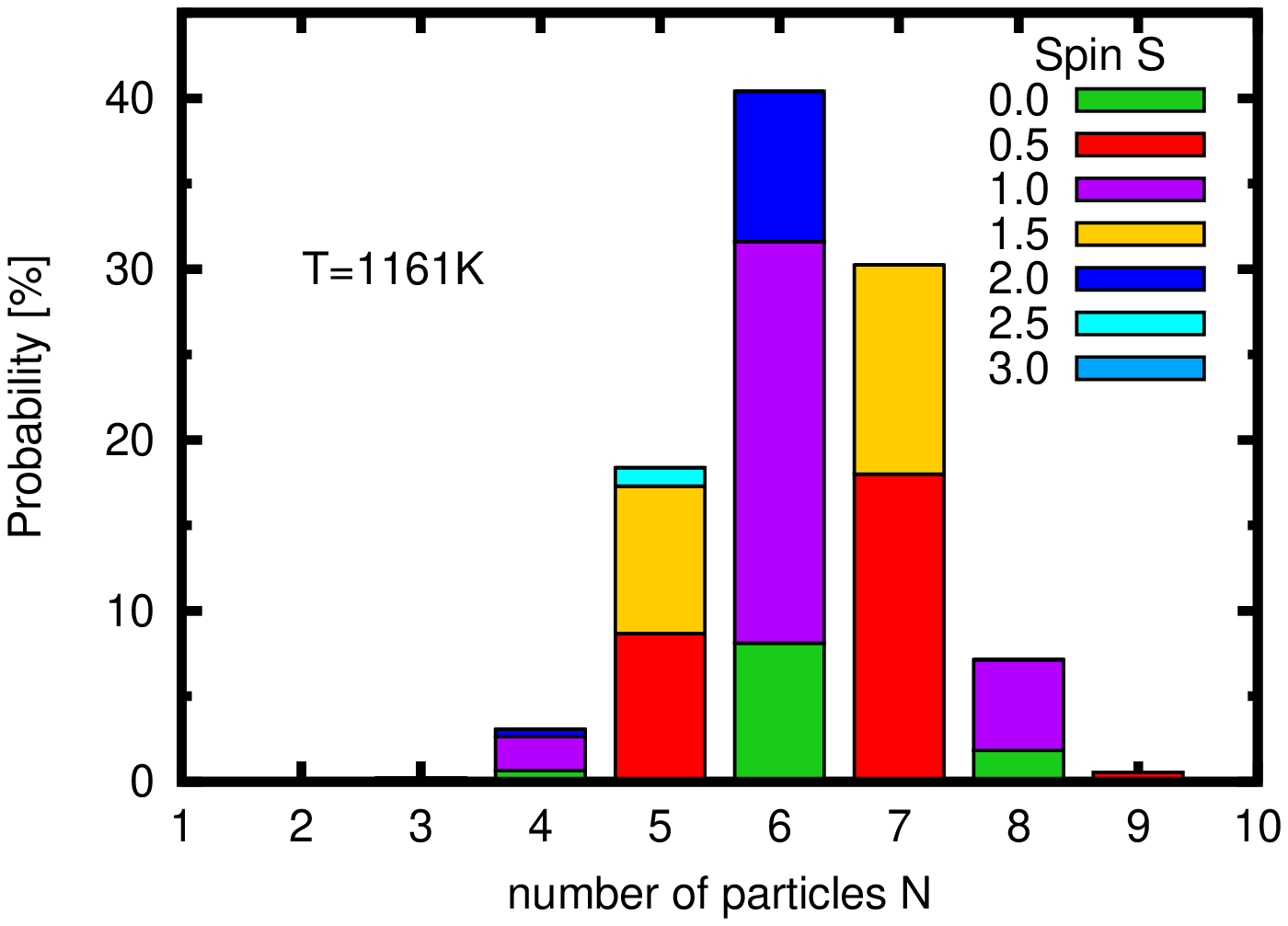}}
      \caption{{\bf The spin and charge state.} Decomposition of the DMFT impurity onto the eigenstates of the effective atom, resolved into the spin $S$ and charge state $N$.}
      \label{fig_histo}
      \end{center}
\end{figure}

Having thus ascribed the metalization process in FeSi to an effect of electronic correlations, we may ask as to the physical origin of this coherence-decoherence crossover.
For this we look at the microscopic insights furnished by our theoretical approach. Within dynamical mean field theory the system is described by an effective impurity 
that represents the iron atoms,
and a hybridization function (Weiss field) that accounts for the embedding of the reference system into the solid.
Useful information can be gained by decomposing the local projection of the system into the eigenstates of the effective impurity, as shown in \fref{fig_histo}.
Displayed is the probability of the local reference system to be in a state with $N$ particles and spin state $S$, for (a) low, and (b) high temperature.

The histogram shows a rather broad distribution over many spin and charge states, signaling large fluctuations at short time scales.
Indeed the variance of the charge state $\delta N=\langle\left( N-\langle N\rangle\right)^2\rangle\approx 0.93$ is of order unity, entailing an overall mixed valence state,
with an average 3d occupation $\left\langle N\right\rangle=6.2$.
Also virtual spin fluctuations are large, $\delta S=\langle\left( S-\langle S\rangle\right)^2\rangle\approx 0.33$.
Further, we obtain from the spin distribution an effective moment $M=\sqrt{S(S+1)}g_s\approx3$ ($g_s=2$), which is consistent with major contributions from $S=1$ states,
and in congruence with the moment
$M=2.7$ as obtained from fitting a Curie-Weiss law $\chi=\frac{\mu_0\mu_B}{3k_B}M^2/(T-T_C)$ to
the experimental susceptibility 
\cite{PhysRev.160.476,JPSJ.50.2539} for $T>400$K.
The dominance of states with $S=1$ implies in particular that FeSi is not a singlet insulator\cite{PhysRevLett.76.1735,vanderMarel1998138}.

Further, we note that the decomposition shown in \fref{fig_histo} (a) \& (b) is basically independent of temperature.
This means that the mixed valence state is not temperature induced, as previously proposed\cite{PhysRevB.50.9952}.
Moreover, this also rules out a spin state transition.
This is to be contrasted to systems like MnO, or LaCoO3 in which high-spin/low-spin transitions occur
\cite{kunes_mno,jmt_mno,PhysRev.155.932,PhysRevLett.106.256401}. 

The temperature independence of the variances of the spin and charge state (short time fluctuations) is in stark contrast to the 
experimental uniform susceptibility (time averaged response) which shows activated behaviour at low T, and a Curie-Weiss like decay at temperatures beyond 400K.
Thus, there is a tremendous differentiation in time-scales: While the underlying spin structure of the (effective) iron atoms does not evolve, the manifestation of the fluctuating moment in the spin response is highly susceptible to external conditions%
\footnote{For the theoretical local spin susceptibility see Ref.\cite{jmt_fesi}.}.

The temperature induced unlocking of the fluctuating moment of the iron sites establishes a link to real space.
Therewith the momentum space description of the low temperature coherent semiconductor breaks down and effective lifetimes are introduced
as the system decomposes over states of different momenta.

\section{The Seebeck coefficient of FeSi and RuSi}
\label{seebeck}

Finally we turn to the discussion of the thermoelectric properties of FeSi.
For details of the employed linear response formalism, see Refs. \cite{jmt_fesb2,jmt_fesi}, and e.g.\ Refs \cite{PhysRevLett.67.3724,PhysRevB.65.075102,oudovenko:035120,JPSJS.71S.288,Held_thermo,Haule_thermo} for related prior works.

\begin{figure}[htp]
\centering
\includegraphics[angle=-90,width=0.95\textwidth]{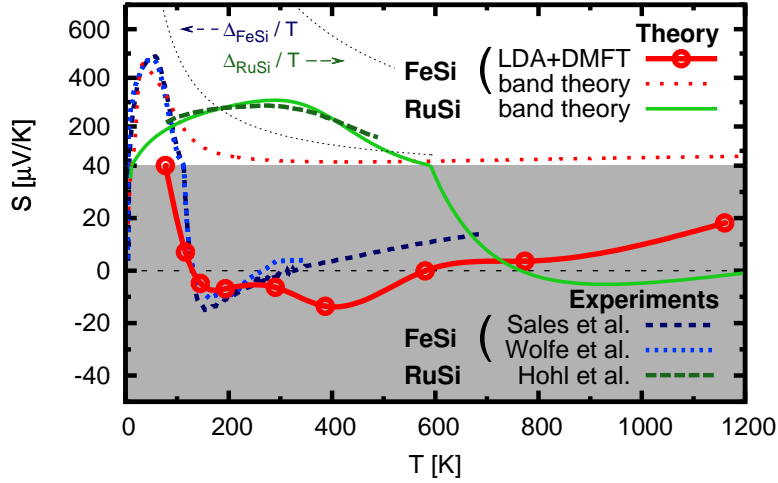}
\caption{{\bf The Seebeck coefficient of FeSi and RuSi.} While a description of FeSi requires employing a many-body theory, the thermopower of RuSi is reproducible
based on band-structure methods, see text for details. Experimental data from \cite{Wolfe1965449,PhysRevB.83.125209} (FeSi) and \cite{Hohl199839} (RuSi). Indicated with thin dotted lines are the maximal electronic contributions to the Seebeck coefficient $\Delta/T$ for both compounds.}
\label{fig_S}
\end{figure}

Before addressing the influence of many-body effects onto the thermopower, 
we note that it was found that --for temperatures below 100K-- the Seebeck coefficient of FeSi can actually be reproduced by 
a slightly hole doped band-structure \cite{PhysRevB.59.15002}. This we confirm by adding 0.001 holes per iron as well
as an overall effective mass of two to the density functional results from above.
The thus obtained Seebeck coefficient indeed yields a good agreement at low temperatures, as shown in \fref{fig_S}. 
In our opinion, the hole doping
should not be viewed as an introduction of extra charge, but as a way to adjust the particle/hole asymmetry
of the band-structure\footnote{However, a notable dependence of the thermopower on the precise stoichiometry is witnessed in experiments\cite{Buschinger1997784}.}.

The congruence of a band-structure based thermopower with experiment further corroborates that FeSi is a coherent semiconductor at low temperatures.
In the regime $k_BT\ll \Delta$ the thermopower of such a system is -- modulo the temperature independent Heikes contribution -- given by $(\Delta/2-\mu)/T\cdot \delta\lambda$,
where $\Delta$ is the charge gap, $\mu$ is the chemical potential measured from the centre of the gap, and $\delta\lambda$ quantifies the electron/hole asymmetry
(for details see Ref. \cite{jmt_fesb2}).
Since the latter is constrained, $\left|\delta\lambda \right|\le 1$ ($\delta\lambda=+1$ would e.g.\ correspond to a purely hole driven thermopower),
the electronic contributions to the Seebeck coefficient are, for an insulator, limited to $\pm\Delta/T$. 
Thus, assuming the same asymmetry, a larger gap causes a greater Seebeck coefficient.
As indicated in \fref{fig_S}, FeSi respects the above boundary, while it is, for example, largely surpassed
in the related compound FeSb$_2$\cite{jmt_fesb2}, advocating for that material the importance of 
non-electronic contributions, especially the phonon-drag mechanism.
The quenching of the Seebeck coefficient at very low temperatures ($S\rightarrow0$ for $T\rightarrow0$) is the consequence of a small yet finite
scattering rate\cite{jmt_fesb2}\footnote{For transport calculations that are based on density functional theory we assume a constant scattering rate $\Im\Sigma=-1$meV.}.

\medskip

The use of a renormalized band-structure fails at describing the properties of FeSi above 100K, when the metalization process becomes notable,
as seen in the above \fref{fig_spec} for the spectral function (photoemission experiments), and now in \fref{fig_S} for the Seebeck coefficient.
Extending our scheme\cite{jmt_fesb2} for the thermopower to include dynamical self-energy effects, we calculate the Seebeck coefficient
based on our realistic dynamical mean field results. As displayed in \fref{fig_S}, the thus obtained thermopower is in very good agreement
with experimental findings. In particular, we capture the changes of sign as a function of temperature which indicate the transition between
hole ($S>0$) and electron ($S<0$) dominated transport. The non-monotonous tendencies in the Seebeck coefficient were already heralded by
the moving of the chemical potential, as seen in the inset of \fref{fig_spec}: Starting from low temperatures, the chemical potential moves down,
therewith reducing the hole contributions to the Seebeck coefficient, before it passes, at around 120K, the point of thermoelectric particle/hole symmetry,
below which the Seebeck coefficient becomes negative. At yet higher temperatures, the trend reverses and the thermopower changes sign again.

Thus, in the current case correlation effects are detrimental for the thermoelectric performance. This has to be contrasted to the case
of correlated metals, where a reduced quasi-particle weight increases the Seebeck coefficient (see e.g.\ Refs.~\cite{PhysRevB.65.075102,Haule_thermo}).

\medskip

We find it instructive to compare FeSi to its iso-structural and iso-electronic homologue RuSi, and
discuss the series Fe$_{1-x}$Ru$_{x}$Si in view of its potential as thermoelectric%
\footnote{For a comparison of the related couple FeSb$_2$ and RuSb$_2$, see Refs. \cite{APEX.2.091102,herzog_fesb2}.}.
Ruthenium silicide, RuSi, is a semiconductor with a gap of $0.2-0.3$eV as inferred from optical spectroscopy\cite{Buschinger1997238,Vescoli1998367}
or resistivity measurements\cite{Buschinger1997238,Hohl199839}. 
Interestingly the tendency in the size of the gap in Fe$_{1-x}$Ru$_{x}$Si is not monotonous in $x$\cite{PhysRevB.65.245206}.
Indeed up to a ruthenium concentration of 6\% the charge gap is found to decrease. With increasing concentration the gap then grows, bypasses
the initial value of FeSi at around 15\% ruthenium, and further augments up to the value of pure RuSi.

With our insight into FeSi, we can explain this trend. As a matter of fact, there are 
two opposing tendencies that have to be considered:
On the one hand, ruthenium has a larger atomic radius than iron.  Thus with increasing Ru content, the hybridization gap (see \sref{bnd}) will shrink as the lattice expands.
This effect is immediate, and wins for low ruthenium concentrations.
On the other hand, one has to consider the crossover in the dominant orbital character of excitations at low energy, namely the transition
from 3d to 4d electrons.
In FeSi we identified the ratio of the Hund's rule coupling $J$ and the bandwidth as the controlling factor for the strength of correlations\cite{jmt_fesi}.
Since the 4d electrons of ruthenium are less localized than the iron 3d ones, this ratio decreases with growing ruthenium admixtures, resulting
in smaller effective masses and thus a weaker many-body narrowing of the charge gap.
This effect becomes preponderant for larger ruthenium concentrations.

This decrease in electronic correlation effects, of course, also means that the use of conventional band-structure methods is more justified for RuSi than for FeSi:
Using again the GGA functional within wien2k\cite{wien2k}, we obtain a band-structure (not shown)
that resembles that of FeSi, albeit with stronger dispersions and a larger gap.
The latter is, as in FeSi, indirect, and amounts to 0.23eV, in agreement with previous
band-structure calculations\cite{Imai2006173,0295-5075-85-4-47005,springerlink:10.1134/S1063782609020031}.
For the Seebeck coefficient, we here find for RuSi -- as was the case for FeSi -- that a small hole doping is needed to adjust the particle/hole asymmetry in 
density functional results.
Indeed, as shown in \fref{fig_S}, doping pure RuSi with only 0.0025 holes per ruthenium, yields a thermopower in excellent agreement with experiment\cite{Hohl199839}.

The Seebeck coefficient of RuSi, while not reaching the very large values of FeSi at low temperatures, is notable in size over an
extended temperature regime, with about $250\mu \hbox{V/K}$ from 100 to 500K.
Comparing RuSi with FeSi, we make two interesting observations:
(a) as noted above, the thermopower of a coherent insulator is controlled by the size of the gap and the particle/hole asymmetry.
RuSi having a larger gap than FeSi, its Seebeck coefficient indeed surpasses the envelope function of FeSi, $\Delta_{\hbox{\small FeSi}}/T$,
for $200\hbox{K}\le T\le 500$K, as indicated in \fref{fig_S}. 
However, it is further away from its own boundary, $\Delta_{\hbox{\small RuSi}}/T$, than is the case for FeSi.
Hence, the particle/hole asymmetry (above called $\delta\lambda$) is smaller in the 4d compound.
(b) Besides the peak value which is smaller in RuSi, the temperature dependence is qualitatively akin.
Indeed, when scaling the temperature with the respective sizes of their low temperature gaps, the curves for FeSi and RuSi are very similar%
\footnote{It would be of great value to have experimental measurements on RuSi up to higher temperatures.}. 

Without actually performing many-body calculations for RuSi, we can expect that also in other experimental observables the 
temperature dependence scales with the ratio of the respective gap values.
In particular the emergence of a fluctuating moment that causes the coherence-decoherence crossover and quenches the Seebeck coefficient in FeSi
will be pushed to higher temperatures.

Therefore, we believe that the iso-valent substitution of iron with ruthenium, Fe$_{1-x}$Ru$_{x}$Si, will provide competitive thermoelectric properties
in the temperature regime of 100 to 250K.
Besides the greater coherence, the loss of thermoelectric particle/hole asymmetry ($\delta\lambda$) when departing from pure FeSi, is partly compensated by the enlarged gap $\Delta$ (for $x\ge 0.15$).
While the power factor $S^2\sigma$, with the dc conductivity $\sigma$, of pure FeSi reaches favourable
40 $\mu$W/(K$^2$cm) at around 60K\cite{jmt_fesi}, we expect a notable improvement over both pure FeSi and pure RuSi
beyond 100K.
Finally, the substitution will also reduce the thermal conductivity $\kappa$, yielding a better figure of merit
$ZT=S^2\sigma T/\kappa$.

\section*{Conclusions}

In conclusion, we presented a new scenario for the intriguing properties of iron silicide FeSi, in which the metalization
process with temperature is driven by correlation induced incoherence that we traced back to the unlocking of fluctuating iron
moments. Using realistic many-body techniques, we investigated the signatures of this microscopic theory in the one-particle spectrum, the optical conductivity (see Ref.\cite{jmt_fesi}), and the Seebeck coefficient, for all of which we find
quantitative agreement with experiment.

With the thus gained physical insight into the interplay of electronic correlation effects and thermoelectricity,
the general class of narrow gap semiconductors becomes more amendable to theory assisted thermoelectric material design.
Indeed, we made explicit suggestions for improvements of FeSi based thermoelectrics.
While pure FeSi already presents a competitive power factor in the temperature range of 50-100K,
we proposed means for extending the favourably large Seebeck coefficient to higher temperatures, by motivating
a study of Fe$_{1-x}$Ru$_{x}$Si vis-{\`a}-vis its thermoelectric performance.

\section*{Acknowledgments}
We thank F. Steglich, P. Sun, and S. Paschen for stimulating discussions. JMT further acknowledges IICAM travel support through the NATO advanced workshop
``The New Materials for Thermoelectric Applications: Theory and Experiment'' in Hvar,
as well as the hospitality at MPI CPfS, Dresden.
The authors were supported by the NSF-materials
world network under grant number NSF DMR 0806937 and NSF DMR 0906943, and by the PUF program.
Acknowledgment is also made to the donors of the American Chemical Society Petroleum Research Fund 48802 for  partial support of this research.

\end{document}